\documentclass[preprint]{aastex631}
\usepackage{epsfig}
\usepackage{graphicx}
\usepackage{longtable}
\usepackage{natbib}
\usepackage{subfigure}
\usepackage{xcolor}
\usepackage{amsmath}
\usepackage{multirow}
\usepackage{todonotes}

\newcommand{\Msolar}{M$_{\odot}$}

\begin{document}
\shorttitle{Sub-subgiants in \textit{Gaia} EDR3}
\shortauthors{Leiner et al.}

\title{Revealing the Field Sub-subgiant Population Using a Catalog of Active Giant Stars and \textit{Gaia} EDR3}

\author{Emily M. Leiner}
\altaffiliation{NSF AAPF Fellow}
\affiliation{Center for Interdisciplinary Exploration and Research in Astrophysics (CIERA) and Department of Physics and Astronomy, Northwestern University, 1800 Sherman Ave., Evanston, IL 60201, USA}
\email{emily.leiner@northwestern.edu}

\author{Aaron M. Geller}
\affiliation{Center for Interdisciplinary Exploration and Research in Astrophysics (CIERA) and Department of Physics and Astronomy, Northwestern University, 1800 Sherman Ave., Evanston, IL 60201, USA}
\affiliation{Adler Planetarium, Department of Astronomy, 1300 S. Lake Shore Drive, Chicago, IL 60605, USA}

\author{Michael A. Gully-Santiago}
\affiliation{Department of Astronomy, The University of Texas at Austin, Austin, TX 78712, USA}

\author{Natalie M. Gosnell}
\affiliation{Department of Physics, Colorado College, 14 E. Cache La Poudre St, Colorado Springs, CO 80903}

%\author[0000-0003-2053-0749]{Benjamin M. Tofflemire}
\author{Benjamin M. Tofflemire}
\altaffiliation{51 Pegasi b Fellow}
\affiliation{Department of Astronomy, The University of Texas at Austin, Austin, TX 78712, USA}

\begin{abstract}
Sub-subgiant stars (SSGs) fall below the subgiant branch and/or red of the giant branch in open and globular clusters, an area of the color-magnitude diagram (CMD) not populated by standard stellar evolution tracks. One hypothesis is that SSGs result from rapid rotation in subgiants or giants due to tidal synchronization in a close binary. The strong magnetic fields generated inhibit convection, which in turn produces large starspots, radius inflation, and lower-than-expected average surface temperatures and luminosities. Here we cross-reference a catalog of active giant binaries (RS CVns) in the field with \textit{Gaia} EDR3. Using the \textit{Gaia} photometry and parallaxes we precisely position the RS CVns in a CMD. We identify stars that fall below a 14 Gyr, metal-rich isochrone as candidate field SSGs. Out of a sample of 1723 RS CVn, we find 448 SSG candidates, a dramatic expansion from the 65 SSGs previously known. Most SSGs have rotation periods of 2-20 days, with the highest SSG fraction found among RS CVn with the shortest periods. The ubiquity of SSGs among this population indicates SSGs are a normal phase in evolution for RS CVn-type systems, not rare by-products of dynamical encounters found only in dense star clusters as some have suggested. We present our catalog of 1723 active giants, including \textit{Gaia} photometry and astrometry, and rotation periods from \textit{TESS} and VSX. This catalog can serve as an important sample to study the impacts of magnetic fields in  evolved stars.
\end{abstract} 

\section{Introduction}
Observations show that rapid rotation in stars and the resulting strong magnetic fields can have significant impacts on the evolution of stars and their positions in the Hertzsprung-Russell (HR) Diagram. Observations of active, rapidly-rotating M-dwarfs and brown dwarfs yield temperatures and radii that significantly deviate from stellar models (e.g. \citealt{Torres2010, Jackson2019, Morales2010}), with rapid rotators showing cooler temperatures and radius inflation. A similar phenomenon is found in pre-main-sequence stars (e.g. \citealt{Hillenbrand1997, Somers2015}) and other active main-sequence stars (e.g. \citealt{OBrien2001}). In some star clusters, observations reveal multiple main-sequence turnoffs, with more rapid rotators forming a redder turnoff than their slower-rotating counterparts \citep{Bastian2009, Sun2019}. It has been hypothesized that the radius inflation and cooler surface temperatures observed in these stars may arise from magnetic inhibition of convection \citep{Chabrier2007, Feiden2013} and/or starspots \citep{Torres2006}, or enhanced mixing and gravity darkening if the star is rotating fast enough to become oblate \citep{Bastian2009}. 

\begin{figure}[htp]
    \centering
    \includegraphics[width= .9\linewidth]{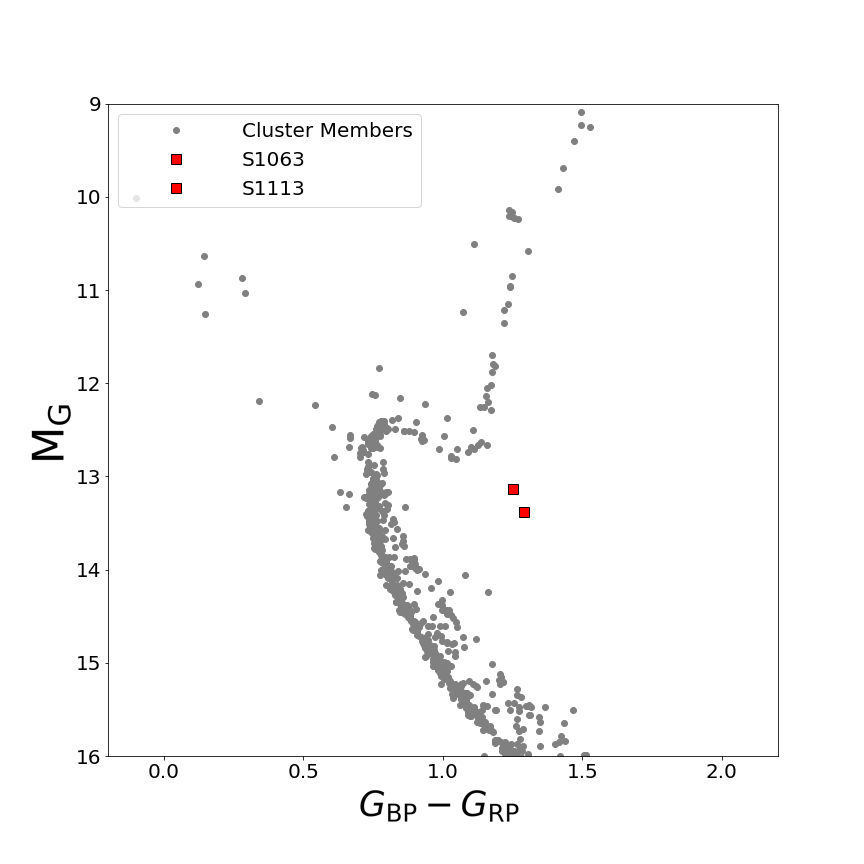}
    \caption{The old (4 Gyr) open cluster M67 hosts two of the best-studied SSGs, S1113 and S1063 \citep{Mathieu2003}. These stars are fainter and redder than the subgiant and giant branch in a region not occupied by standard stellar evolutionary tracks. All \textit{Gaia} astrometric members of the cluster are shown with gray points, S1063 and S1113 are shown with red squares. Membership selection is done as described in \citet{Leiner2021}.}
    \label{fig:m67_cmd}
\end{figure}

Similar deviations from standard models in temperature, radius, and luminosity may occur in subgiant and giant stars that are rapidly rotating due to tidal synchronization in a close binary. Sub-subgiant stars (SSGs) are found in color-magnitude diagrams of both open and globular clusters fainter than the subgiant branch and/or redder than the giant branch (Figure~\ref{fig:m67_cmd}; \citealt{Geller2017, Mathieu2003}), an area difficult to explain given standard stellar evolution tracks or the combined light of normal cluster stars. \citet{Geller2017} present a catalog of $65$ SSGs found in open and globular clusters. These SSGs
share several characteristics. They are generally X-ray sources, photometric variables, H-$\alpha$ emitters, and where their binary status is known, they are typically found to be binaries with orbital periods of 2-18 days \citep{Geller2017}. These properties indicate SSGs are magnetically active binaries with giant or subgiant primary stars. As such, they may be an under-luminous category of the general class of active giant binaries known as RS Canum Venaticorum (RS CVn) stars \citep{Hall1976}, which are not generally understood to be under-luminous. In RS CVn systems, tidal synchronization spins up the rotation of the primary, yielding a rapidly rotating and magnetically active giant star. These RS CVn giants have large starspots which produce photometric variability of up to several tenths of a magnitude, X-ray emission, and spectroscopic indicators of chromospheric activity like H$\alpha$ or Ca II H \& K emission. We note that \citet{Geller2017} use the term ``sub-subgiant" to refer specifically to stars less luminous than the subgiant branch, and ``red straggler" to refer to stars that fall to the red of the giant branch, and their sample of 65 sources includes both. We do not make this distinction here, and refer to stars both fainter and redder than the giant branch all as SSGs. 

Due to their magnetic properties, \citet{Leiner2017} hypothesize SSGs are subgiants or lower red giant branch (RGB) stars that have lower observed temperatures and luminosities because their magnetic fields inhibit convection and generate large starspots. Spots are inferred from the photometric variability observed in most SSGs, and in one SSG in M67 a spot covering fraction varying from 20$-$45\% has been measured using spectral decomposition and light curve analysis \citep{Gosnell2021}. Alternatively, the authors suggest that SSGs could be subgiant or giant stars losing mass via Roche lobe overflow in a close binary, or the products of dynamical collisions between stars that occur in dense cluster environments. \citet{Geller2017b} conclude that Roche lobe overflow or collisions would only rarely produce SSGs. On the other hand, the magnetic fields hypothesis might produce SSGs frequently enough to explain their observed numbers in clusters.   However, the stellar physics behind this mechanism is difficult to fully incorporate into stellar evolution models and therefore the impact on stellar evolution is still not well understood.  Importantly, this magnetic fields hypothesis does not require cluster dynamics (as does stellar collision), and therefore can operate just as effectively within the galactic field.

No dedicated search for SSGs has yet been undertaken in the field, where identifying SSGs is more challenging than in clusters with a clear isochrone. However, some authors have identified possible candidates in the field (i.e. the No Mans Land stars in \citealt{Huber2014} or the active subgiant star identified in \citealt{Tucker2021}). Here we seek to understand if SSGs are found outside of clusters among the galactic field population and to explore the connection between the SSGs and the RS CVn stars. Are all or most RS CVn stars found to be under-luminous like the SSGs, or are SSGs a small subset of all magnetically active giant stars? And if this is the case, what determines whether an RS CVn will be underluminous rather than falling on a standard isochrone? 

To do this we use a sample of known RS CVn stars from past and ongoing variability surveys contained in the VSX catalog \citep{Watson2006}. We use parallax measurements from \textit{Gaia} EDR3 to confirm their classification as giant stars and to position them precisely within a color-magnitude diagram, as well as to perform other quality cuts to ensure we are using a sample with low extinction and reliable photometry and astrometry (Section~\ref{section:sample}). 
We assemble rotation period measurements for this large sample from the VSX catalog as well as the ongoing \textit{TESS} mission (Section~\ref{section:rotation}). From this sample, we identify those stars that appear to be fainter and/or redder than a normal subgiant or giant, and categorize these stars as candidate SSGs (Section~\ref{section:SSGselection}). We use this sample of SSGs to discuss the general properties of the SSG population and what this may tell us about the origins of these unusual stars (Section~\ref{section:discussion}). 

\section{Observations and Sample}
\label{section:sample}
\subsection{VSX RS CVn Sample}
To begin, we take the entire sample of active giants stars (RS CVn stars) variables identified in the VSX catalog \citep{Watson2006}.  We use the April 2021 version of this catalog. This database first consisted of the Combined General Catalog of Variables stars and the New Catalog of Suspected Variables (NSV), as well as variable star discoveries from the literature.  It has since been updated to include variability information from many recent variability surveys including the Northern Sky Variability Survey (NSVS; \citealt{Wozniak2004}), 3rd All Sky Automated Survey (ASAS-3; \citealt{Pojmanski2002}), the Optical Gravitational Lensing Experiment (OGLE-II; \citealt{Udalski1997}), Zwicky Transient Facility (ZTF; \citealt{Masci2019}), Catalina Sky Survey (CSS; \citealt{Drake2009}) and other variability surveys and literature sources. The catalog is therefore not a homogenous sample of stars observed in the same photometric bandpasses and limiting magnitudes, but it does offer a large selection of variable stars observed across the entire sky. We select from the VSX catalog all stars designated \texttt{RS} for RS CVn, which gives us 73,203 RS CVn candidates as a starting point for our sample. 

\subsection{\textit{Gaia} EDR3 distances and photometry}
\begin{figure*}%[!phbt]
\centering
\includegraphics[angle=0, width= .45\linewidth ]{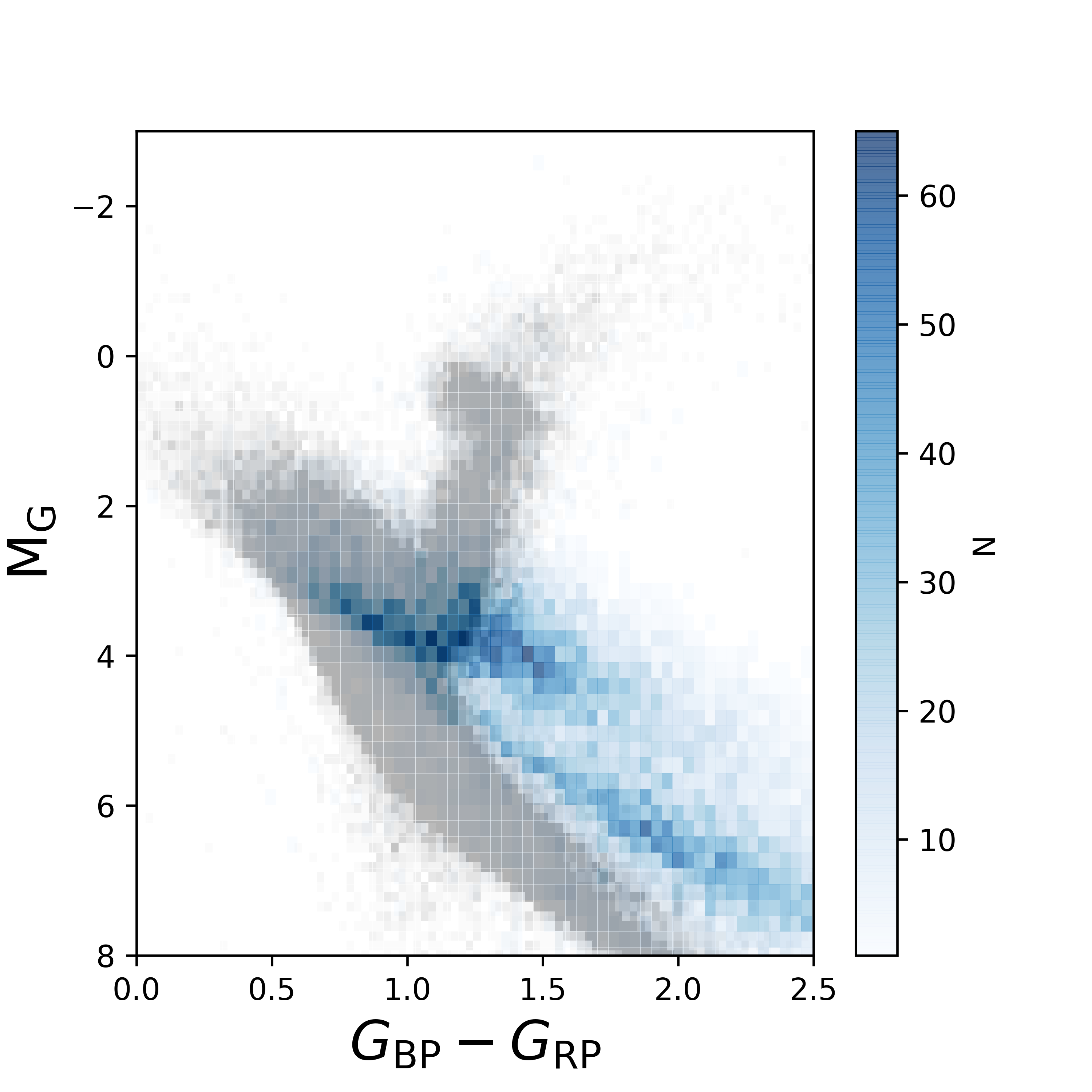}
\includegraphics[angle=0, width= .45\linewidth ]{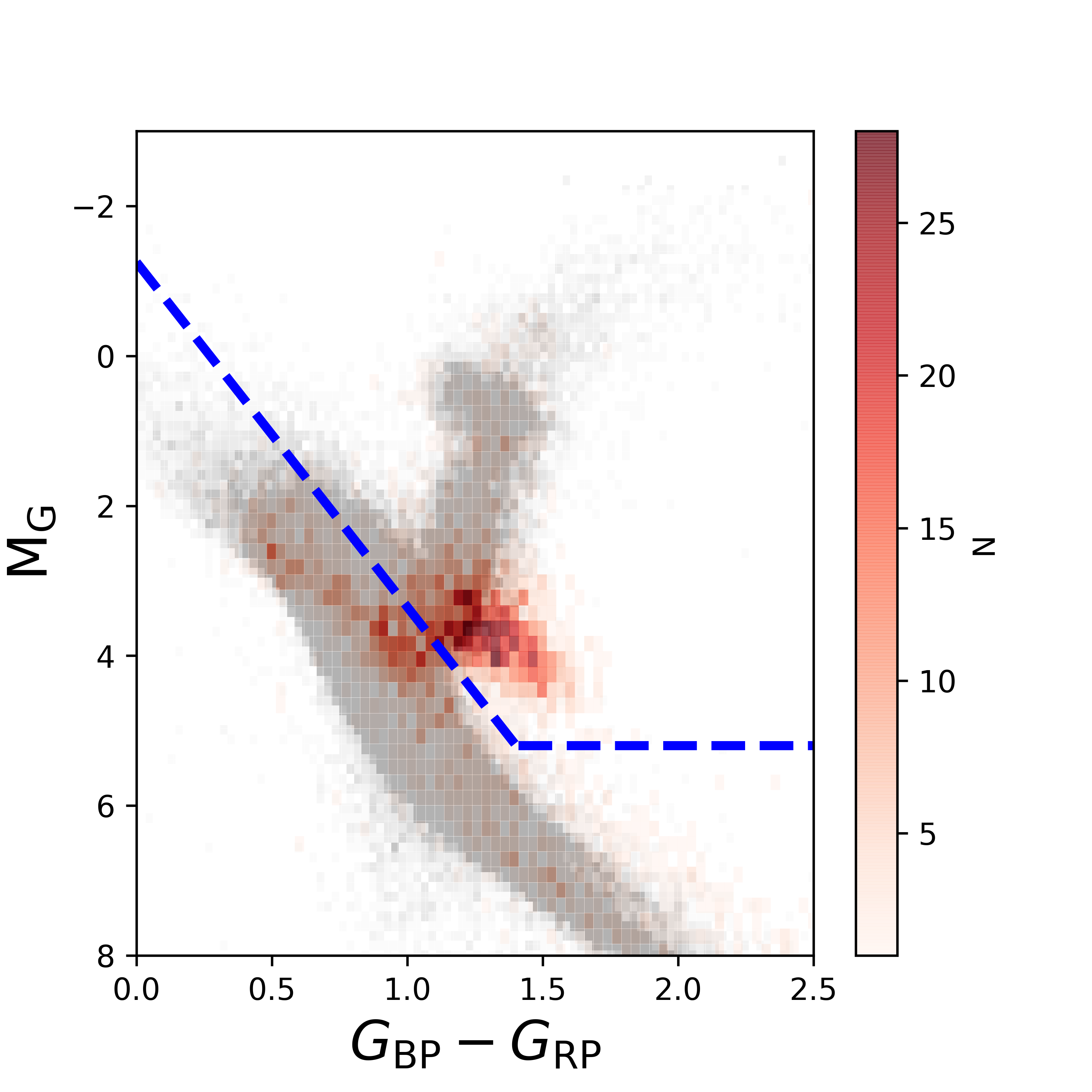}
\caption{We show a 2D histogram of the full sample of RS CVn candidates within 2000 pc  for which we find matches in the \textit{Gaia} catalog (left), and the subset of this sample that has A$_V< 0.5$ (right). Some misclassified main-sequence variables remain in the low-extinction sample. We therefore remove any stars with photometry placing them below the blue line in this CMD. In both plots we also show a density plot of stars in the Kepler field within 2000 pc for comparison (gray).}
\label{fig:samplecmd}
\end{figure*}

We match this sample of 73,203 VSX RS CVns to the \textit{Gaia} EDR3 catalog \citep{GaiaEDR3} based on their RA/Dec coordinates. Specifically, we query the \textit{Gaia} EDR3 database for sources within $1\farcs4$ of each VSX source, selecting those sources with only one match in the catalog within this radius. This approach works for the majority of stars (70,902 sources), but we find that it excludes many of the brightest, closest sources in the catalog which may have higher proper motions. We therefore take any unmatched stars after this first attempt and look for nearby targets within progressively larger radii out to $12\farcs$. When we find the nearest neighbor, we check that the $G-$band magnitude is within half a magnitude of VSX magnitude (generally $V-$band or similar). Using this approach we find an additional 559 matches. In total, then, we find unique \textit{Gaia} matches for 71,463 sources in the VSX catalog. 

For these 71,463 sources we extract parallax measurements from the \textit{Gaia} catalog and invert them to determine distance. We exclude any stars with distances larger than 2000 pc to further ensure the stars in our sample have good distance solutions and accurate photometry. We compare these distances to the probabilistic geometric distances in \citet{Bailer-Jones2021}, which include the zeropoint correction from \citet{Lindegren2021}, and find that they differ by at most a few percent for stars in our relatively nearby sample. Our choice of systematic parallax corrections therefore does not significantly impact our conclusions, and we use the standard inverted parallax throughout this work. In addition to these distances, we adopt photometry from the \textit{Gaia} catalog in $G$, $G_\mathrm{BP}$, and $G_\mathrm{RP}$ bandpasses so that we have uniform photometry across the entire sample. These cuts yield a sample of 22,372 stars (Figure~\ref{fig:samplecmd}, left panel).

%or these Gaia stars from the Bailer-Jones catalog \citep{Bailer-Jones2018}, which uses a Bayesian approach to determining distances to Gaia stars using their parallaxes. We follow their suggestion to exclude stars with a bimodal distance solution or a poorly constrained distance (catalog flags \texttt{modality flag} and \texttt{result flag}). We exclude any stars with distances larger than 2000 pc to further ensure the stars in our sample have good distance solutions and accurate photometry. In addition to these distances, we adopt photometry from the Gaia catalog in $g$, $bp$, and $rp$ bandpasses so that we have uniform photometry across the entire sample. These cuts yield a sample of 22,372 stars.  

\subsection{Extinction and Color-Magnitude cuts \label{section:cuts}}
The VSX-\textit{Gaia} sample of 22,372 stars described above clearly includes a substantial number of stars that are not RS CVn, as they fall on or near the main sequence (Figure~\ref{fig:samplecmd}; left panel). These interlopers likely appear because spotted main sequence stars, T Tauri variables, and RS CVns all have variability originating from rotation and star spots. Discerning between these variability classes based on the light curve alone is difficult, and so many of these T Tauri or main sequence variables end up misclassified. 

\begin{figure}[htp]
\centering
\includegraphics[angle=0, width= .99\linewidth, trim={0cm 5cm 0 6cm}]{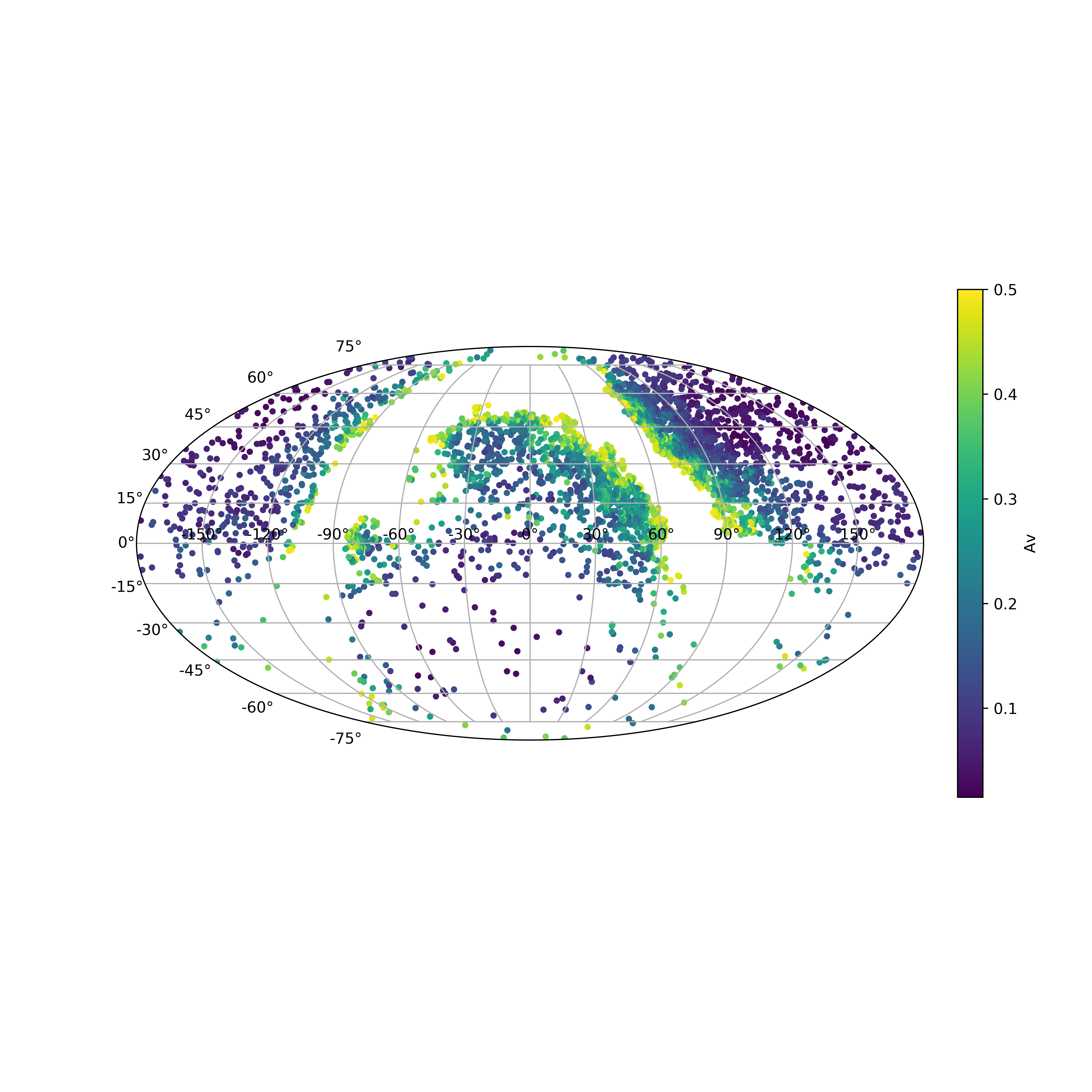}
\caption{We show a Mollweide projection map (RA/Dec.) showing line-of-sight extinction to the VSX RS CVn sample within 2000 pc. The color shows the value of $A_V$ from low (purple) to high (yellow). Sources with $A_V > 0.5$ are excluded from the sample, which removes many high-extinction sources near the galactic plane. }\label{fig:extinction}
\end{figure}

As a first cut to remove T Tauri variables, we calculate line-of-sight extinction values ($A_V$) using the \texttt{vespa} python package \citep{Morton2012}. This package relies on extinction maps from the NASA/IPAC Extragalactic Database, producing extinction-at-infinity values given an object's coordinates. The results show that our RS CVn sample has a broad distribution with some targets extending to high extinction values around $A_V \approx 1.5$. The high reddening tail is likely to contain many weak lined T Tauri stars, which are found in star forming regions near the galactic plane, as well as other highly extincted targets that are difficult to photometrically correct. We therefore impose a cut in $A_V$, including only low-extinction targets with $A_V < 0.5$. This cut has the dual purposes of excluding most of the likely T Tauri stars, as well as removing objects whose CMD location may be significantly influenced by dust extinction/reddening. This cut results in a sample of 3729 stars and removes most stars above the main sequence, a region expected to be occupied by T Tauri stars (see Figure~\ref{fig:samplecmd}, right panel). We show a projection map of the stars in our remaining sample in Figure~\ref{fig:extinction}, color-coded by $A_V$. This cut removes stars that fall in the highly extincted region near the galactic plane. 

The \textit{Gaia} distance and photometry indicate that still some stars remain in the sample that are main sequence stars, not RS CVn giants (Figure~\ref{fig:samplecmd}; right panel). We therefore impose a cut in color and magnitude, removing stars that fall below the blue dashed line in the right panel of Figure~\ref{fig:samplecmd}, thus removing all the stars on or near the main-sequence. This reduces are sample to 1895 stars. 

As a final data quality assurance measure, we reject astrometric solutions with poor astrometric goodness of fit (large \texttt{$\chi^2_\mathrm{AL}$}) or fewer than 9 good observations (\texttt{$N_\mathrm{AL} < 9$}), cuts proposed by \citet{Arenou2018} (see their Equation 1). We do not use photometric filtering as \citet{Arenou2018} suggest, because these stars do have unusual colors/magnitudes and therefore could be filtered out. We also reject sources with a Renormalized Unit Weight Error (\texttt{RUWE}) greater than 1.4, as values much larger than 1.0 indicate single-star astrometric models do not yield a good astrometric fit for these sources. These are likely binary star systems or higher order multiples in which the system is unresolved or partially resolved (e.g. \citealt{Belokurov2020}). For bright variable stars, large amplitude magnitude changes may also inflate RUWE \citep{Belokurov2020}. This effect is small at the magnitudes and amplitudes of typical RS CVn, and we remove only a small number of sources from our sample based on RUWE, so we do not expect this to be problematic for our sample selection. This series of cuts removes 172 sources, nearly all due to high \texttt{RUWE}. 

Out of our original list we therefore select in total of 1723 stars that have well-determined \textit{Gaia} matches, astrometric solutions, and low extinction values, that appear to be true RS CVns rather than misclassified MS stars. In Table~\ref{table:bigtable}, we present a catalog of these stars including \textit{Gaia} EDR3 source ids, VSX catalog IDs, RA, Dec, \textit{Gaia} photometry (G, G$_\mathrm{BP}$, and G$_\mathrm{RP}$), distances, extinction values, and the astrometric quality indicators \texttt{RUWE}, \texttt{$\chi^2_\mathrm{AL}$}, and \texttt{$N_\mathrm{AL}$}. We also provide information on photometric variability, including VSX rotation periods and \textit{TESS} rotation periods, light curve amplitudes, and available sectors of data. We discuss this in detail in Section ~\ref{section:rotation}. The full machine-readable Table~\ref{table:bigtable} is available electronically.

%and rotation periods, which we discuss further in Section~\ref{section:rotation}. We show the first few lines of Table 1 in this paper to illustrate the contents, and the full table is available electronically. 

In addition to the 1723 sources we use in our final sample, we provide in Table~\ref{table:bigtable} the 172 sources we removed from our sample due to astrometric quality cuts. While these sources are not suitable for our investigation here due to their larger astrometric uncertainties, this sample probably contains true RS CVn including many binaries that may be appropriate for follow up, particularly if subsequent \textit{Gaia} data releases provide better astrometric solutions for these sources. We categorize these 172 sources as ``cut" in the final column of Table~\ref{table:bigtable}.

\section{Rotation Periods}
\label{section:rotation}

\subsection{VSX Periods}
The VSX catalog contains rotation periods for all the stars in our sample measured from many previous and ongoing variability surveys (see Section~\ref{section:cuts}). We include source names from the VSX catalog (which indicates the heritage of all measurements) as well as the measured rotation periods from this catalog in Table 1. We do not include light curve amplitudes, as we note that the bandpass used varies between surveys and so these amplitudes are not directly comparable across sources. 

In Figures~\ref{fig:PGplot}-\ref{fig:CMD_4panel} we investigate the RS CVn sample with respect to the VSX rotation periods.  We discuss these plots further in Section~\ref{section:discussion}. 

\subsection{\textit{TESS} periods}

Many of these sources have also been recently observed with \textit{TESS}. We search the \textit{TESS} archive for observations of every star in our sample, finding light curves are available for 1101 of the 1723 RS CVn stars. The 30-minute cadence of the \textit{TESS} observations is much faster than what was used for most periods derived in the VSX catalog, and they have the advantage of all being observed in the same \textit{TESS} bandpass. However, each \textit{TESS} sector is only about 27 days long, and most sources have only one sector of data, so the VSX light curves generally have a much longer time baseline. Because of incomplete sample coverage and shorter time baseline of \textit{TESS} light curves, we can only obtain \textit{TESS} rotation periods for about half of our sample here and therefore we do not include \textit{TESS} rotation periods in our population analysis. However, the higher-precision \textit{TESS} light curves may be a valuable resource to follow-up studies that can yield insights into spot properties and possible binary companions (\emph{i.e.} if some sources are eclipsing binaries). We therefore match our sources to the \textit{TESS} archive and provide TIC numbers and preliminary rotation periods for sources where it is possible. We provide this information in Table 1 so the information is easily available for follow-up studies.

To determine preliminary \textit{TESS} rotation periods we use the software package \texttt{lightkurve} \citep{2018ascl.soft12013L} to download \textit{TESS} full frame images (FFIs) and process the photometry. We use coordinates from the \textit{Gaia} catalog to select pixels from the \textit{TESS} FFIs, using a cutout size of 5 pixels around the given coordinate. We use \texttt{lightkurve} functions to perform aperture photometry,  defining target pixels as those containing flux $3\sigma$ above the median level, and background pixels to be those with flux $<0.1\sigma$ above median levels. We then subtract the mean background flux per pixel from the target lightcurve, normalize the lightcurve, and use sigma-clipping to remove outliers greater than 5~$\sigma$ from the mean photometric value. We again use \texttt{lightkurve} to calculate a periodogram for each light curve using an oversample factor of 10, taking the period of maximum power in this periodogram as the star's rotation period. We calculate the root mean squared amplitude of the variability by subtracting the $25^{th}$ percentile flux of the lightcurve from the $75^{th}$ percentile and dividing by $\sqrt{2}$. We report both this period and amplitude in Table~\ref{table:bigtable} where available. 

For most stars in our sample, only 1 \textit{TESS} sector is available, and the period and amplitude we derive is based only on this sector. We did not attempt to stitch lightcurves for sources possessing multiple sectors. Such stitching can hypothetically access those stars exhibiting modulation on timescales longer than 27 days, but in practice the sector-to-sector offsets and other long-term trends hamper their automated and unbiased analyses.  Instead, we simply repeat our light curve analysis for each available sector independently and report the median value of the period and amplitude in our table. We also show in Table~\ref{table:bigtable} how many sectors of data were used when determining this median value. In some cases not all sectors were used in the period determination because of data quality issues, or because the star fell too close to the edge of the detector. A higher quality period may be determined for longer period sources with multiple \textit{TESS} sectors by carefully stitching different sectors of data together, but we leave this non-trivial work to future authors. 

Taking only the sources with VSX periods less than the 27 days , we find agreement to within 10\% between the \textit{TESS} and VSX rotation periods for $\sim50$\% of these sources. In cases where the two periods do not match within 10\%, the causes seem to be varied. Due to the large \textit{TESS} pixel size, targets with bright nearby stars in the field may have substantial flux contamination and require more careful background subtraction to recover their periods. This is particularly true for the fainter stars in our sample. If we limit the sample to those stars brighter than $G= 13$, we find over $70\%$ of these sources have consistent periods.  In other cases, the periods measured are within a factor of 2 of each other, suggesting that harmonic aliasing causes the discrepancy. VSX periods measured near 1 day also are confirmed by the \textit{TESS} observations only $\sim5\%$ of the time. In these cases the ground-based periods are likely to be spurious detections. In other cases, the \textit{TESS} light curve simply does not show significant variability, perhaps due to an error in the initial VSX period determination, a change in activity level between the VSX observations and the \textit{TESS} observations, or a very low amplitude variation that requires more specialized light curve processing to uncover. In still other cases, systematics in the \textit{TESS} light curve such as long term trends or warming events due to data downloads in the middle of the \textit{TESS} sector introduce erroneous periodicities. In short, a star-by-star analysis is needed to understand the source of discrepancy for each source individually, and to manually tune the light curve analysis parameters to find the correct period if possible. We leave this more detailed analysis to future work. 

We conclude that the VSX periods are generally in agreement with \textit{TESS} periods for bright, short-period sources where TESS data is reliable, with the exception of the VSX periods measured near 1 day which are more likely spurious. TESS periods are more likely to be unreliable for faint sources in our table ($G> 13$) and longer period sources (particularly those with $P_\mathrm{rot} > 27$ days, the length of a typical TESS sector). We adopt the VSX periods for the remainder of our analysis. 

\section{SSG Selection}
\label{section:SSGselection}

\begin{figure*}
    \centering
    \includegraphics[width= 0.95\linewidth]{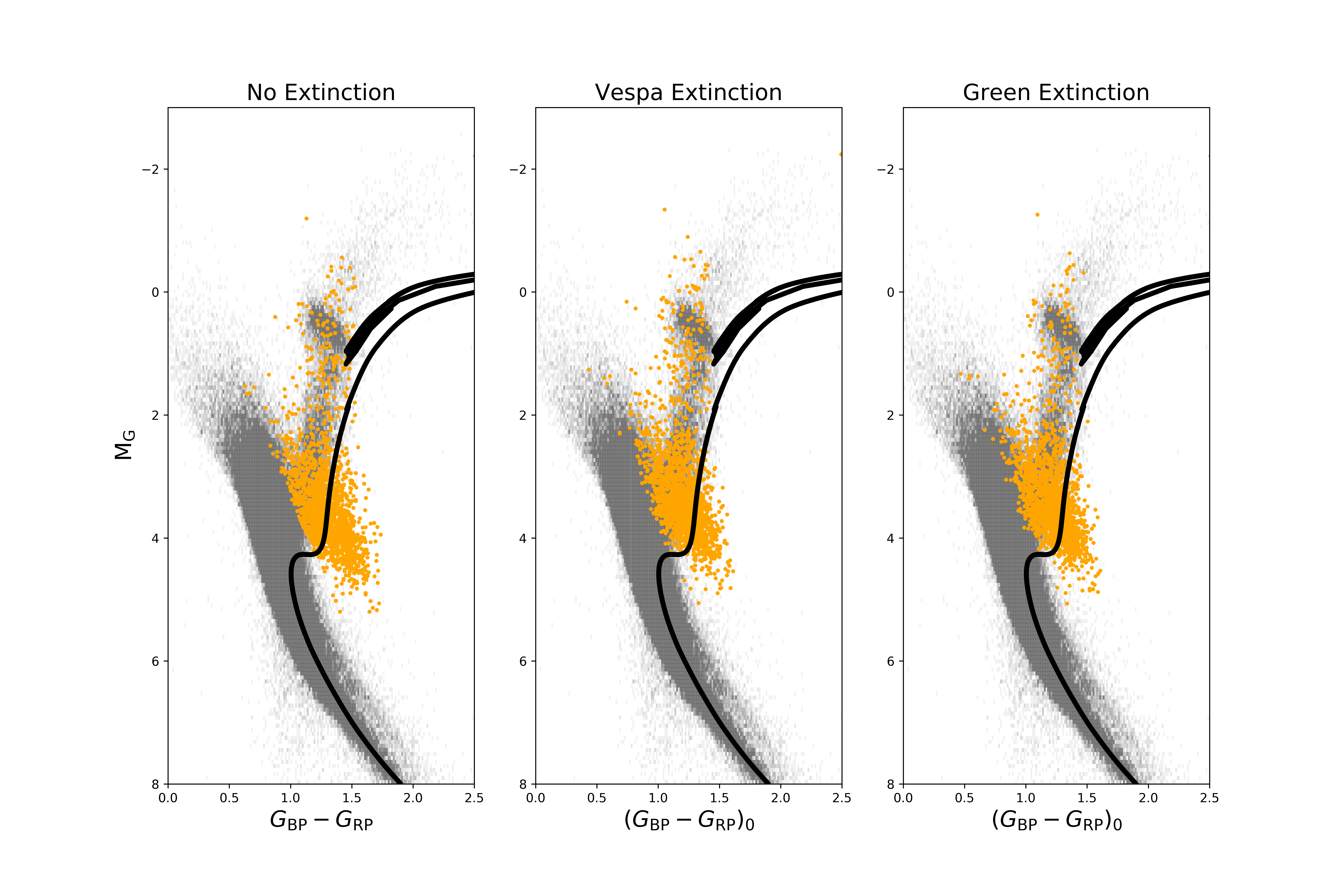}
    \caption{We show CMDs of the RS CVn sample using various reddening/extinction corrections. On the left, we show the sample with no reddening correction applied. In the middle we use line-of-sight extinctions from \texttt{vespa}. On the right we use 3D extinction corrections from \citet{Green2019}. We also show the Kepler sample within 2 kpc in gray, and a 14 Gyr, [Fe/H]= +0.5 MIST isochrone in black.\label{fig:cmd_period_panels}}
\end{figure*}

In Figure~\ref{fig:cmd_period_panels}, we show the 1723 RS CVn stars in our final sample in orange. For reference, we also show a 14 Gyr, [Fe/H] = +0.5 MIST isochrone \citep{Dotter2016}. This isochrone defines the red edge of the giant domain, e.g., the location of the oldest, most metal rich, and therefore coolest red giants we would expect to exist in the field. We find that a significant fraction of the RS CVn sample is fainter and/or redder than this isochrone; we define these as candidate SSG stars in the field.  This is a conservative cut, as it requires a star to have a color and magnitude redder and fainter than even the most metal-rich, low-mass giants expected given the age of the universe. 

Also in Figure~\ref{fig:cmd_period_panels} we investigate the impact of our extinction correction on the identification of SSGs.  First, we show the CMD with no reddening/extinction corrections applied to the sample (left panel). We also show the CMD with extinction corrections applied to the entire sample using \texttt{vespa} line-of-sight extinctions (center panel). Finally, we use the 3D extinction map of \citet{Green2019} to correct the sample using a distance-dependent extinction (right panel). This map is only available for Dec. $> -30.0$ degrees, so for this plot we eliminated sources from our sample at lower declinations. Our sample contains few sources in the southern sky (Figure~\ref{fig:extinction}), so this cut removes $< 5$\% of our sample. We use the extinction law of \citet{Wang2019} to convert the extinction map values to \textit{Gaia} photometric bandpasses.

We find that without any extinction correction, 46\% of the RS CVn sample are fainter and/or redder than the 14 Gyr metal rich isochrone, and therefore reside within the SSG region.  Using extinctions-at-infinity from \texttt{vespa}, 26\% of the sample reside in this region. Using the 3D map of \citet{Green2019} results in 28\% of the sample reside in the SSG region. We conclude from this that using line-of-sight extinctions rather than distance dependent extinctions does not result in significant changes to our results. Since the \texttt{vespa} map covers the entire sky, we use these extinctions for the remainder of our work. We report the classification of each star in our sample as either an SSG or RS CVn in the final column of Table~\ref{table:bigtable}. 

SSGs make up more than a quarter of our RS CVn population in the field. Furthermore, the true fraction of RS CVn that are SSGs could be significantly higher because our selection criteria is biased towards detecting only lower mass SSGs, and we do not account for each star's mass, age, and metallicity in our RS CVn sample. For example, a 2.0 \Msolar\ solar-metallicity giant star could be a very under-luminous SSG before it would appear as an SSG given our cut, whereas we would expect to select most 1.0 \Msolar, metal-rich SSGs given our criteria. We return to this point in the next section.

\section{Discussion} \label{section:discussion}
The catalog of RS CVn and SSGs that we present above gives us the largest sample of SSGs to-date and provides new insights into the properties of the overall SSG population. Below we discuss the observed frequencies, magnitudes, CMD distributions, and rotation periods of the RS CVn sample as a whole and the SSG sub-sample. We also discuss what these observations may reveal about the formation mechanisms of the SSGs.  
\subsection{Magnitudes and Rotation Periods}
\label{section:SSGperiods}
\subsubsection{RS CVn}
In Figure~\ref{fig:PGplot} we show the distribution of rotation periods and absolute magnitudes of the entire RS CVn sample (gray points) as well as the subset of the sample classified as SSGs (red points). The RS CVn sample as a whole extends approximately from $M_G= 5$ to $M_G= -1$ in absolute magnitude, and displays rotation periods ranging from approximately 1 to 100 days. However, the main locus of RS CVn extends from rotation periods of $\sim2-20$ days and $4.5 >M_G> 2.5$.  The overdensity of points near $P= 1$ days is likely due to the spurious ground-based VSX periods that are not confirmed by TESS (see Section~\ref{section:rotation}). 

Furthermore, we notice that the RS CVn sample covers a fairly distinct diagonal region on this plot, extending from shorter periods at fainter absolute magnitudes to longer periods at brighter absolute magnitudes.  We suggest that these bounds may be set at brighter magnitudes and larger radii by the Roche radius, and at fainter magnitudes and smaller radii by the interplay between the the radius where tidal synchronization is sufficiently effective to spin up the stars, and the limits of magnetic saturation (which decreases for longer periods and larger radii).   

First we investigate the bright end of the sample.  Assuming that binary orbits are circular and the orbital periods and rotation periods are synchronized in these systems, at a given period we can calculate the maximum size a star could reach before exceeding its Roche lobe and beginning mass transfer. To do this, we use the equation for Roche lobe radius from \citet{Eggleton1983}, which depends on the mass ratio of the binary system and the orbital separation of the binary. We assume that stellar radius is equal to this Roche radius in order for Roche lobe overflow to begin. We then identify the point on a MIST evolutionary track \citep{Dotter2016} where a star of the given mass and metallicity reaches this radius. We can then infer an limit in absolute magnitude by finding the $M_G$ that corresponds to this point on the MIST evolutionary track. We show this limit with the blue and black MIST evolutionary tracks \citep{Dotter2016} in Figure~\ref{fig:PGplot}, varying our assumptions about primary mass, secondary mass, and metallicity. The upper bound of our distribution agrees well with this limit if the RS CVn are predominately 1-2 \Msolar\ stars of near solar-metallicity, reasonable values for a nearby sample of red giant stars. 

The morphology at the faint end of our sample appears to be more complex. For periods up to $\sim10-15$ days, the minimum absolute magnitude in our sample is $M_ G\sim5$. However, at rotation periods longer than this, the minimum magnitude steadily increases. By a rotation period of $\sim30$ days, the minimum magnitude has decreased to $M_G\sim3$. 

Tidal synchronization is the mechanism by which RS CVn are spun up to the rapid rotation velocities needed to generate strong magnetic fields and large starspots, and tidal forces drop off quickly with wider binary separations. An orbital period near 10 days marks the tidal circularization cutoff period in solar-type main sequence binaries (e.g. \citealt{Geller2021, Lurie2017, Meibom2005}). Such binaries with periods less than 10 days are found in tidally synchronized and circularized orbits.  At slightly larger orbital periods these binaries transition to non-circular pseudo-synchronized systems, and the widest binaries are beyond the reach of tides.  Synchronized systems wider than 30 day orbits are rarely found among solar-type binaries \citep{Lurie2017}. A similar limit of $P \lesssim 30$ days may be applicable to the lowest luminosity (smallest radius) RS CVns, as they will not have grown much bigger than their main-sequence size. For these long-period and low luminosity systems, tidal forces may not be strong enough to synchronize rotation and orbital periods.  At longer periods, only larger (and therefore brighter) RS CVns will have been spun up by tidal synchronization. Without tidal spin-up, red giant binaries will be rotating too slowly to generate the strong magnetic fields required for classification as an RS CVn. 

Additionally, other studies have found that at rotation periods of 10-15 days, low-mass main-sequence stars undergo a transition in magnetic activity between a saturated and unsaturated regime. At this point, indicators of magnetic activity such as X-ray luminosity or H-$\alpha$ emission begin to decrease (e.g. \citealt{Noyes1984, Wright2018}), and magnetic activity begins to decrease linearly with slower rotation periods. \citet{Dixon2020} find a similar decrease in near-UV emission (another indicator of magnetic activity) among red giant stars, with a drop-off in activity observed at $\frac{P_\mathrm{orb}}{\sin{i}}= 10$ days. This transition may also be mass dependent; a study of intermediate-mass giants ($1.5 < M < 3.8$ \Msolar) finds X-ray emission begins to drop off at rotation periods closer to $P_\mathrm{rot} =100$ days \citep{Gondoin2007}, though this is based on a small sample size. If true, this mass dependence would explain the existence of longer-period magnetically active stars at the brightest magnitudes but not among the fainter sources. 

In short, these observations suggest that the mechanism that produces strong magnetic fields and starspot activity (leading to RS CVn classification) is most effective at short periods and perhaps also mass dependent, such that less massive giants must have shorter periods than the more massive giants to display signs of strong magnetic activity. 
%with longer rotation periods do not generate the strong magnetic fields and starspot activity necessary for classification as an RS CVn for the fainter, smaller giants, but larger and/or more massive giant stars can be quite active at these rotation periods. 

Additionally, we note that longer rotation periods are more difficult to detect photometrically due to the longer time baseline of observations required. Observational incompleteness may therefore be significant among long-period systems, though given the long time baseline of many of the VSX observations and the detection of many systems with periods as long as 100 days, it is unlikely this limiting absolute magnitude is simply an observational bias. 

\subsubsection{SSGs}
\begin{figure}
    \centering
    \includegraphics[width= .98\linewidth]{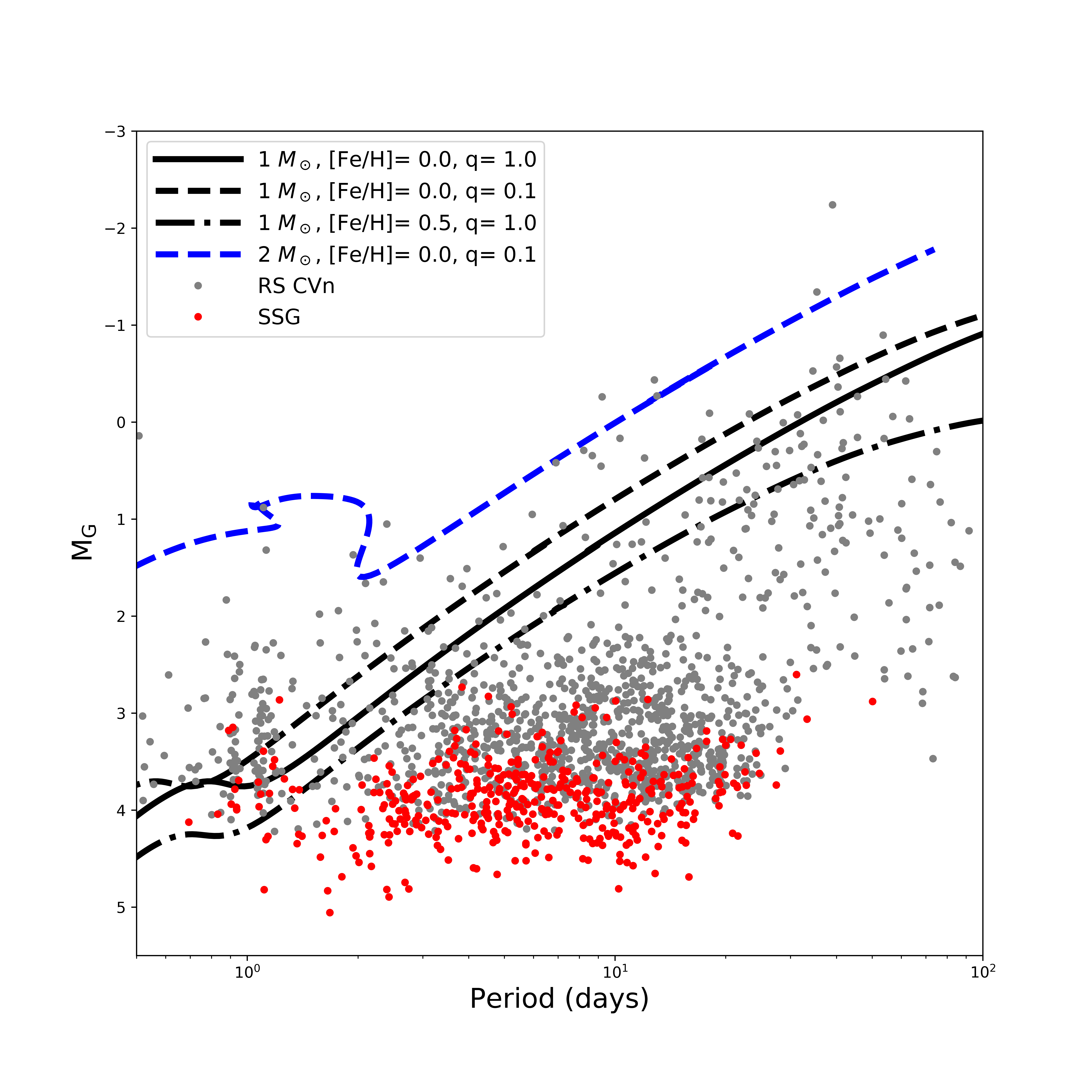}
    \caption{We show the extinction-corrected \textit{Gaia} G-band absolute magnitude (M$_\mathrm{G}$) as a function of the rotation period for the RS CVns (gray points) and SSGs (red points) in our sample. We also show the maximum magnitude expected as a function of orbital period for various MIST stellar models (black and blue lines) assuming that the star is completely filling its Roche lobe.}
    \label{fig:PGplot}
\end{figure}

In Figure~\ref{fig:PGplot}, we highlight stars categorized as SSGs with red points. These SSGs span a G-band absolute magnitude range from $3<M_G< 5$. SSGs with orbital periods longer than 20 days are uncommon, making up just 5\% of the SSG population, and we find only 1 of 448 SSGs with an orbital period longer than 35 days. 

\begin{figure*}
    \centering
    \includegraphics[width= 0.48\linewidth]{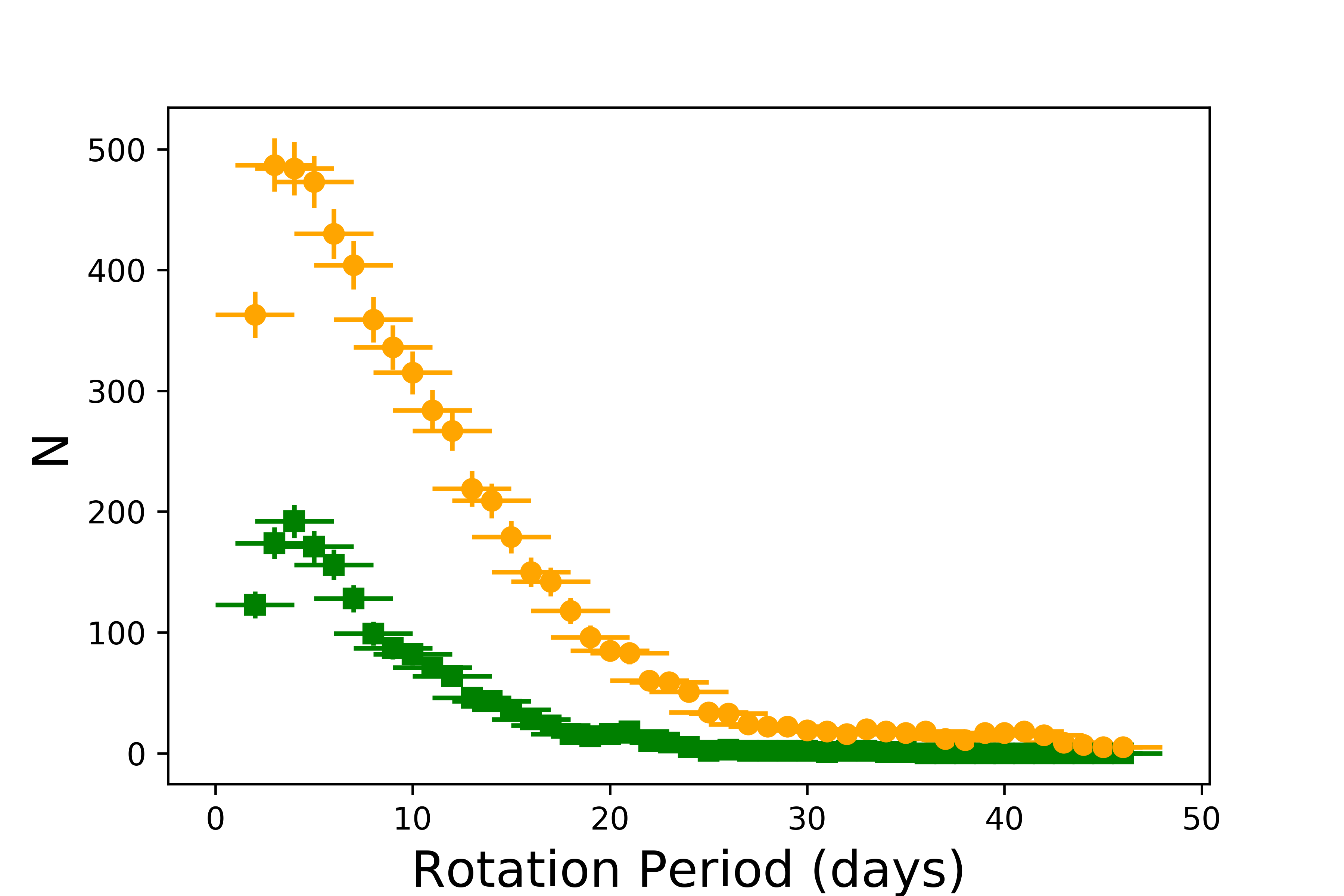}
    \includegraphics[width= 0.48\linewidth]{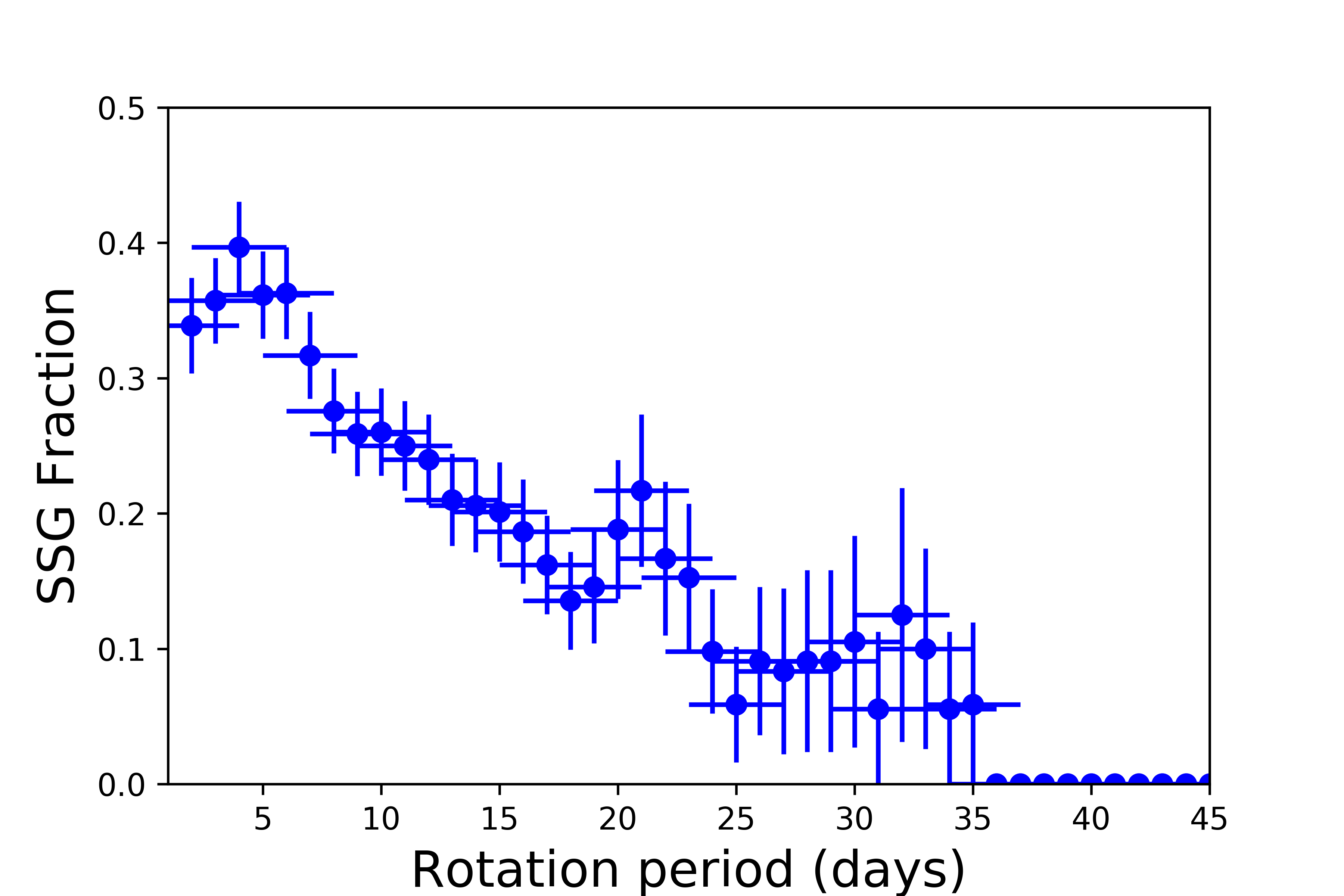}
    \caption{(left)We show the number of SSGs (green) and total RS CVn (orange) in our sample as a function of the VSX rotation period. Error bars indicate the Poisson errors on $N_\mathrm{SSG}$ and the size of the period window used ($\pm 2$ days).  (right) We show the fraction of SSGs in each period bin. That is, we divide the number of SSGs in each period bin by the total number of RS CVn in our sample within the same bin. \label{fig:SSG_fraction}}
\end{figure*}

Shown another way, in Figure~\ref{fig:SSG_fraction} we plot the number of total RS CVn in our sample (left panel, orange) and SSGs in the sample (left panel, green) as a function of the rotation period from VSX. We calculate total number within a sliding period window of $\pm 2$ days. The SSG period distribution is similar to that of the full RS CVn sample, peaking at rotation periods around 5 days before declining. On the right, we also show the fraction of SSGs as a function of period (i.e., number of SSGs divided by number of RS CVn in each period window). At the 5-day period peak of the distribution, about 40\% of the RS CVn sample is categorized as an SSG. The fraction then declines, leveling to about $20\%$ at periods of 10-20 days. Beyond 20 days, the SSG fraction declines quickly and we find only 1 SSG in our sample with a rotation period $> 35$ days. This decline is due to the decreasing absolute magnitudes of the RS CVn sample at longer rotation periods (Figure~\ref{fig:PGplot}) that we discuss above, coupled with a decline in the the total number of RS CVn with increasing rotation periods. 

These findings are consistent with \citet{Geller2017}, who find that cluster SSGs have periods ranging from 2-18 days. While active giants with longer rotation periods exist, they appear to be found exclusively among larger and brighter giants rather than the SSG population. SSGs have a similar rotation period distribution to the RS CVn overall, suggesting the SSGs can be understood as simply the faint end of the RS CVn population.

%Across all periods, 27\% of the sample meets this criteria for a SSG star, but this SSG fraction varies by period (Figure~\ref{fig:SSG_fraction}). For stars with $P<5$ days, 35\% are SSGs. For stars with $5 \leq P < 10$, 32\% are SSGs. For stars with $10 \leq P < 20$, 24\% are SSGs. For stars with $P\geq 20$, only 10\% are SSGs. In Figure~\ref{fig:SSG_fraction}, we show a plot of the number of total RS CVn in our sample (left panel, orange) and SSGs in the sample (left panel, green) as a function of the rotation period from VSX. On the right, we also show the fraction of SSGs as a function of period (i.e., number of SSGs divided by number of RS CVn in each period bin). The SSG period distribution is  similar to that of the full RS CVn sample, though the SSG distribution is slightly shifted towards shorter rotation periods. 

%Figure~\ref{fig:SSG_fraction} shows that the overall number of both RS CVn and SSG peaks near rotation periods of 5 days, followed by a steady decline and leveling off to close to zero at rotation periods larger than 25 days. The SSG fraction follows a similar distribution overall that peaks near 40\% at 5 days before declining. However, this decline levels off to just above 20\% at 12-20 days. There is another decline between 20-25 days, with the SSG fraction leveling off near 10\% out to 35 days. Beyond 35 days, the SSG fraction drops to zero. 

%Overall this sample indicates that SSGs usually have rotation periods approximately between 2 and 20 days (95\% of the sample), with some SSGs observed with periods between 20-35 days (5\% of the sample), and no SSGs observed with periods greater than 35 days. This is consistent with the finding of \citet{Geller2017} that cluster SSGs have periods ranging from 2-18 days. SSGs follow a similar period distribution to the RS CVn overall, although they are somewhat more common among shorter period systems, and rarely observed among the longest period RS CVn population. As we note previously, our selection criteria biases us towards lower mass SSGs. One possible explanation for this distinction is that longer period RS CVn have higher-mass primary stars, which we discuss in the next section.  
%This would be compatible with the brighter magnitudes observed for the longest period RS CVn in our sample (see the yellow points in Figure~\ref{fig:CMD_4panel}, bottom right panel). 

\subsection{The SSG CMD Region}

\begin{figure*}
    \centering
    \includegraphics[width= 0.95\linewidth]{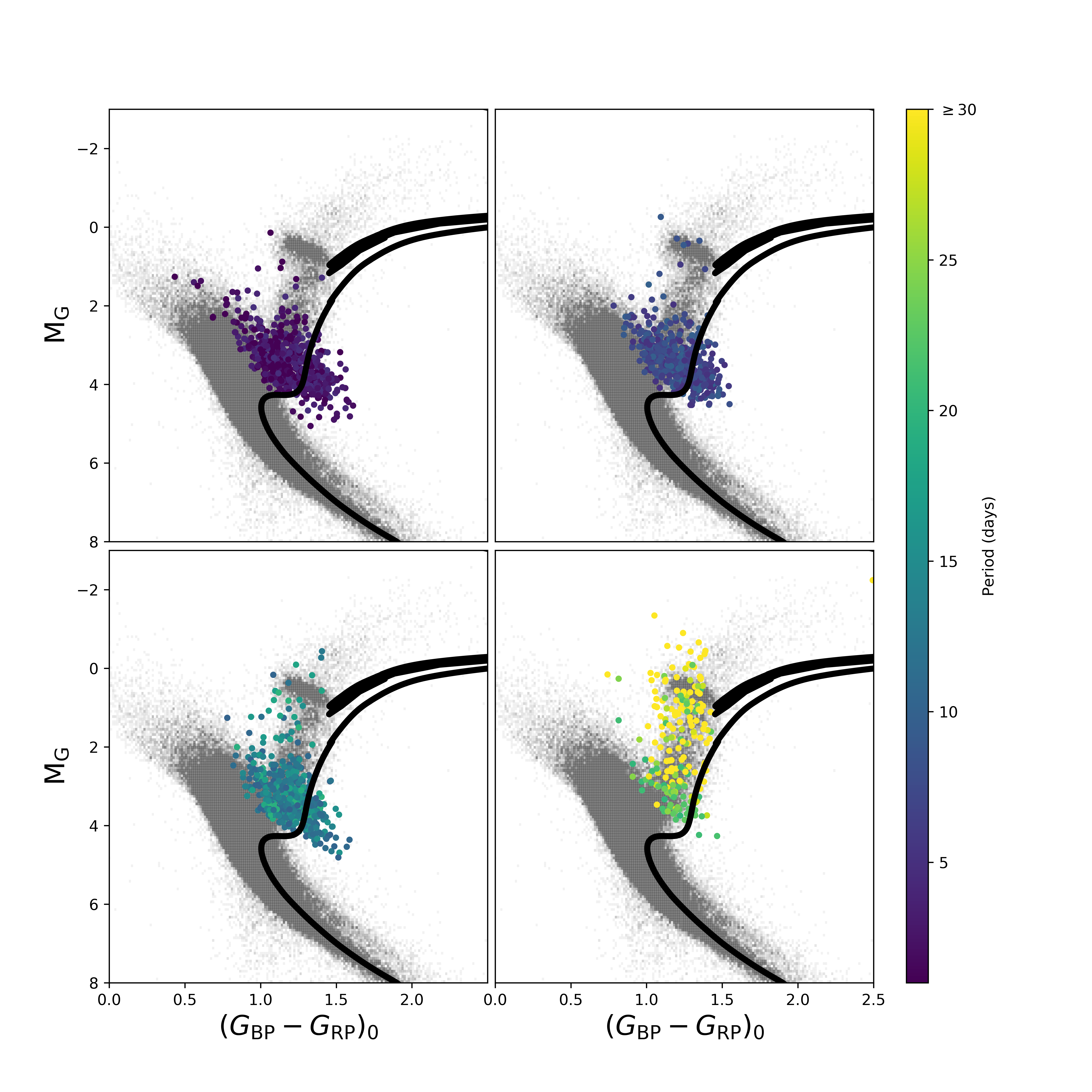}
    \caption{We show the same RS CVn sample as in Figure ~\ref{fig:cmd_period_panels}, but split into four panels by observed VSX rotation period. In the upper left we show stars with $P \leq 5$ days. In the upper right we show $5 < P \leq 10$ days. In the lower left we show $10 < P \leq 20$ days. In the lower right we show $P > 20$ days. We use the \texttt{vespa} extinction corrections. Points are colored by rotation period as indicated by the color bar.} 
    \label{fig:CMD_4panel}
\end{figure*}

In Figure~\ref{fig:CMD_4panel} we show our sample split into 4 different bins based on the rotation period in the VSX catalog. We show periods less than 5 days on the upper left, 5-10 day rotation periods on the upper right, 10-20 day periods on the lower left, and slow rotators with $>20$ day rotation periods on the lower right. 

We find that the SSGs in our sample span from our 14 Gyr isochrone out to $(G_\mathrm{BP}-G_\mathrm{RP})_0 \approx 1.7$. The brightest SSGs extend to $M_G= 3$, and the faintest at  $M_G= 5$.  As we observed in Figure~\ref{fig:PGplot}, the SSGs span a similar range in $M_G$ regardless of rotation period until $P\approx 20$ days, at which point SSGs rapidly disappear from the color magnitude diagram. The spread in $(G_\mathrm{BP}-G_\mathrm{RP})_0$ is also similar across rotation period until the SSGs disappear from the CMD at $P\approx20$ days. 

We clearly see in Figure~\ref{fig:CMD_4panel} that the longer period RS CVn (bottom right panel) fall almost exclusively above the 14 Gyr isochrone, and therefore are not categorized as SSGs in our sample. The longest period RS CVn ($P> 30$ days; yellow points in the bottom right panel of Figure~\ref{fig:CMD_4panel}) are much brighter than the short period RS CVn and fall well to the blue of the 14 Gyr isochrone. Their CMD position indicates these giants are more evolved and/or higher mass giants than those that fall within the SSG region. 

As we note in Section~\ref{section:sample}, if these longer-period rotators are all more massive giants, we would not expect to categorize them as SSGs even if they were not following standard evolutionary tracks. A red giant much larger than 1.0 \Msolar\ would have to be extremely red and/or under-luminous to fall below the 14 Gyr isochrone we use for SSG selection.  Determinations of masses and metallicities for this population are therefore necessary to confirm if this is the reason for the lack of long-period systems in the SSG region. 

\subsection{SSG Formation Mechanisms}
If we make the assumption that rotation periods and orbital periods are synchronized for the RS CVn sample, as we do in the models we show in Figure~\ref{fig:PGplot}, it appears that most SSGs should be well within their Roche lobes. At the shortest periods of just a few days, the $M_G$ magnitudes of the brightest SSGs indicate they may be approaching their Roche radii if the primaries are near solar mass (see Figure~\ref{fig:PGplot}). However, at longer periods ($P \gtrsim 5$ days), the vast majority of both the SSGs and the RS CVn sample as a whole do not appear bright enough to be nearing or exceeding their Roche lobes. If SSGs were produced predominantly via Roche lobe overflow (one of the hypotheses explored in \citealt{Leiner2017}), we might expect SSGs to be larger and thus brighter with increasing period. This may be true for only the brightest RS CVn in the sample, but not for the main locus of the RV CVn overall nor for the SSG population. This sample therefore supports the conclusion from \citet{Leiner2017} that mass transfer would only be expected in the shortest periods SSG systems ($P_\mathrm{rot}\sim1-2$ days), but not for the the majority of the SSG population.

\citet{Leiner2017} also hypothesize that SSGs could form dynamically during collisions, either when a subgiant or giant has part of its envelope removed during a grazing encounter or when two main sequence stars collide. \citet{Leiner2017} and \citet{Geller2017b} find that such collisions happen quite rarely in all but the densest globular clusters, and therefore cannot explain the observed frequencies of SSGs in clusters, especially in the open clusters. Our sample of field SSGs solidifies the conclusions that dynamical encounters are unlikely to form the majority of SSGs. Such encounters would be exceedingly rare in the field, yet we observe that SSGs are common amongst this population of active field binaries. 
%Instead, Figure~\ref{fig:PGplot} indicates that SSGs (and the main locus of RS CVn overall) span a similar magnitude range regardless of rotation period between 1-20 days, and thus that the SSG phase occurs at a similar evolutionary stage across the sample and is not strongly dependent on the rotation (or orbital) period. 

The observed features of this sample are therefore most compatible with the hypothesis from \citet{Leiner2017} that the SSG phase arises from a peak in magnetic activity during subgiant and early giant evolution rather than collisions or mass transfer in a binary. In this case, a drop off in magnetic activity and/or the efficiency of tidal synchronization at periods of $\sim20$ days may explain the rapid decrease in the number of SSGs around this period as we discuss in Section~\ref{section:SSGperiods}. Active giants with rotation periods longer than $\sim20$ days are found almost exclusively among the brightest RS CVn in our sample, perhaps indicating only larger and/or more massive giants have the stellar activity to form large starspots at these rotation periods (Figure~\ref{fig:PGplot}). In summary, magnetic fields seem most plausible to explain the observed numbers and distributions of SSGs in this field sample, and this hypothesis merits further investigation. 

%If, for example, SSGs were produced via Roche lobe overflow as \citet{Leiner2017} suggest, one would expect that the 20 day period Roche-lobe-filling stars were about 2.5 times larger than the 5-day Roche-lobe filling stars, resulting in a difference in brightness of $\sim2$ magnitudes. The observation that these stars occupy a similar magnitude domain is therefore more compatible with the idea that the SSG phase arises from magnetic activity, not mass loss due to the onset of mass transfer in a binary. 

\section{Summary} 
Using \textit{Gaia} EDR3 and rotation periods measured from the VSX catalog, we assemble a sample of $1723$ active giant binaries in the galactic field. Of these active giants, 448 fall below a 14 Gyr, metal-rich isochrone, demonstrating that they are cooler and less luminous than expected from standard stellar evolution models and can therefore be classified as field SSG stars. This sample is a dramatic expansion of the known SSG population, nearly 7 times larger than the previous known sample of $< 70$ stars found only in open or globular clusters. 

From this population, we observe that SSG stars make up $\sim30\%$ of the population of active giant binaries (RS CVn) in our sample, with the fraction being higher for those RS CVn with the shortest rotation periods and lower for the longest rotation periods. We find $95\%$ of the SSGs in our sample have rotation periods between $<20$ days, and only 1 SSG is observed to have a rotation period longer than 35 days. This is consistent  with the $2-18$ day rotation periods observed for SSGs in star clusters \citep{Geller2017}. We suggest that this period limit may stem from some combination of the cessation of tidal synchronization in longer-period binaries and/or a decrease in magnetic saturation at longer rotation periods. 

We find that the RS CVn with rotation periods $2<P_\mathrm{rot}<20$ days occupy a similar region of the CMD regardless of orbital period that extends in absolute magnitude from $2<M_G< 5$. However, the longest period RS CVn ($P_\mathrm{rot} \gtrsim 30$) tend to be brighter ($0< M_G< 3$). We suggest that these slower rotators may be more massive giants than the SSGs, and therefore that more massive giants can maintain tidal synchronization and magnetic saturation at significantly longer periods. If slow rotating RS CVn are found exclusively among more massive giants, they could also still be quite under-luminous without falling below the 14 Gyr isochrone we use to determine SSG status, therefore explaining the paucity of SSGs with rotation periods longer than 20 days. 

The large number of SSGs observed in the field solidifies the conclusion of \citet{Leiner2017} and \citet{Geller2017b} that SSGs are not the rare outcomes of dynamical interactions in dense star clusters, but instead a standard outcome of giant evolution in close binaries. The range of observed rotation periods and extent within the color-magnitude diagram support the conclusion from studies of SSGs in clusters \citep{Geller2017b, Leiner2017} that the most likely origin of these stars is due to luminosity changes from starspots and magnetic activity rather than mass loss due to Roche lobe overflow. 

This sample provides a large population of $\sim1700$ bright RS CVn stars spread across the sky. It is the first large sample to specifically identify SSG candidates in the galactic field, expanding the known sample from just 65 SSG candidates in star clusters to over 500 observed in both clusters and the field. The sample raises many questions, such as what sets the boundaries of the SSG CMD region and the dramatic cutoff in SSG rotation periods around $P\sim20$ days. These observational features provide important clues about the impact of magnetic fields on red giants and will require additional modeling to thoroughly understand. This sample will therefore serve as an important resource for understanding the impact of magnetic fields and rotation on red giant evolution by providing a large sample of bright, nearby test cases that can be observed and modeled in great detail.   

\bibliography{SSG}
\bibliographystyle{aasjournal}

\begin{acknowledgements}
EML is supported by an NSF Astronomy and Astrophysics Postdoctoral Fellowship under award AST-1801937. NMG is a Cottrell Scholar receiving support from the Research Corporation for Science Advancement under grant ID 27528. 

%ADS
This research has made use of the NASA Astrophysics Data System. Some of the data presented in this paper were obtained from the Mikulski Archive for Space Telescopes (MAST). STScI is operated by the Association of Universities for Research in Astronomy, Inc., under NASA contract NAS5-26555. This paper includes data collected by the TESS mission. Funding for the \textit{TESS} mission is provided by the NASA's Science Mission Directorate. This research has made use of the International Variable Star Index (VSX) database, operated at AAVSO, Cambridge, Massachusetts, USA.This work has made use of data from the European Space Agency (ESA) mission {\it Gaia} (\url{https://www.cosmos.esa.int/gaia}), processed by the {\it Gaia} Data Processing and Analysis Consortium (DPAC, \url{https://www.cosmos.esa.int/web/gaia/dpac/consortium}). Funding for the DPAC has been provided by national institutions, in particular the institutions participating in the {\it Gaia} Multilateral Agreement. This research made use of Lightkurve, a Python package for Kepler and \textit{TESS} data analysis \citep{2018ascl.soft12013L}.

\software{Astropy \citep{astropy:2013, astropy:2018}
  matplotlib \citep{hunter07},
  numpy \citep{2020NumPy-Array},
  scipy \citep{2020SciPy-NMeth},
  lightkurve \citep{2018ascl.soft12013L}}

\end{acknowledgements}

\newpage
\begin{rotatetable}
\movetableright=1mm
\begin{deluxetable*}{lllrrrrcccrrrcccc}
%\centerwidetable
\tablecolumns{17}
\tabletypesize{\tiny}

\tablecaption{RS CVn Sample\label{table:bigtable}}
\tablehead{
\colhead{Source ID} & \colhead{VSX ID} & \colhead{TIC} & \colhead{RA} & \colhead{Dec.} &  \colhead{dist.} & \colhead{G} & \colhead{$G_\mathrm{BP}-G_\mathrm{RP}$}& \colhead{RUWE} &\colhead{$\chi^2_\mathrm{AL}$} & \colhead{$N_\mathrm{AL}$} & \colhead{$P_\mathrm{VSX}$}  & \colhead{$P_\mathrm{TESS}$} & \colhead{Amp} &  \colhead{$N_\mathrm{sec}$} & \colhead{$A_V$} & \colhead{Class} \\
\colhead{} & \colhead{} & \colhead{} & \colhead{(deg.)} & \colhead{(deg.)} &  \colhead{(pcs)} & \colhead{(mag.)} & \colhead{(mag.)} & \colhead{} & \colhead{} & \colhead{} & \colhead{(days)}  & \colhead{(days)} & \colhead{} & \colhead{} &  \colhead{(mag.)}  & \colhead{} \\ 
}
\startdata
2448158953886445696 & ZTF J000020.00-030630.5 & 138662652 & 0.08336 & -3.10850 & 1489 & 14.84 & 1.28 & 1.12 & 407.47 & 324 & 6.13 & -- & -- & 0 & 0.11 & RS\\
2874226134920416512 & CSS\_J000131.5+324913 & 83957002 & 0.38142 & 32.82031 & 1238 & 14.71 & 1.36 & 1.0 & 400.64 & 399 & 13.05 & -- & -- & 0 & 0.14 & SSG\\
385028910256998528 & ZTF J000139.60+442609.4 & 259193712 & 0.41504 & 44.43595 & 694 & 12.83 & 1.11 & 0.95 & 620.98 & 442 & 3.31 & -- & -- & 0 & 0.27 & RS\\
2753192410330966912 & ZTF J000146.98+093917.0 & 408505366 & 0.44579 & 9.65473 & 1942 & 15.09 & 1.51 & 1.03 & 370.28 & 351 & 22.03 & -- & -- & 0 & 0.4 & SSG\\
2745167586059626240 & ZTF J000216.58+055915.0 & 403022015 & 0.56911 & 5.98752 & 732 & 11.98 & 1.22 & 1.42 & 1924.19 & 548 & 6.46 & -- & -- & 0 & 0.12 & cut\\
2881005998495586304 & ZTF J000311.16+385513.2 & 194142000 & 0.79654 & 38.92036 & 1292 & 13.7 & 1.34 & 1.04 & 771.15 & 730 & 1.45 & 2.94 & 0.13 & 1 & 0.28 & RS\\
384060304935385984 & GSC 02789-01374 & 194203492 & 1.15196 & 41.40175 & 1024 & 12.37 & 1.08 & 1.08 & 804.1 & 461 & 5.68 & 5.68 & 0.15 & 1 & 0.23 & RS\\
2745853857410858496 & ZTF J000436.80+072117.9 & -- & 1.15334 & 7.35498 & 1724 & 15.47 & 1.45 & 0.98 & 387.25 & 376 & 2.14 & -- & -- & 0 & 0.16 & SSG\\
393379013875213696 & ZTF J000445.27+482238.8 & 201625229 & 1.18866 & 48.37745 & 1269 & 14.42 & 1.12 & 1.56 & 1126.84 & 502 & 0.26 & 0.26 & 0.16 & 1 & 0.31 & cut\\
393941036121609344 & ZTF J000510.00+501158.6 & 201628447 & 1.29169 & 50.19962 & 1784 & 15.53 & 1.35 & 0.96 & 601.53 & 635 & 9.97 & -- & -- & 0 & 0.39 & RS\\
2444348733779214976 & BC Psc & -- & 1.33392 & -5.70761 & 43 & 4.32 & 1.22 & 3.45 & 74244.23 & 239 & 72.93 & -- & -- & 0 & 0.1 & cut\\
2741569502977245568 & ASAS J000606+0345.1 & 293221023 & 1.52546 & 3.75192 & 1742 & 12.62 & 1.25 & 1.01 & 423.66 & 293 & 53.9 & -- & -- & 0 & 0.06 & RS\\
2798131031105732480 & ZTF J000707.70+194737.8 & 301787909 & 1.78209 & 19.79386 & 1449 & 13.87 & 1.49 & 0.97 & 356.06 & 392 & 7.74 & -- & -- & 0 & 0.09 & SSG\\
386309154108776064 & ZTF J000723.64+443218.1 & 439964553 & 1.84852 & 44.53837 & 1840 & 14.8 & 1.3 & 0.98 & 360.07 & 372 & 1.07 & 13.88 & 0.01 & 1 & 0.24 & RS\\
2880679649700656128 & GSC 02781-01010 & 194257167 & 1.88559 & 38.11628 & 792 & 11.55 & 1.38 & 2.08 & 6831.02 & 634 & 41.61 & 19.21 & 0.03 & 1 & 0.27 & cut\\
2849876625289185536 & ZTF J000808.57+243951.6 & 427721991 & 2.03571 & 24.66435 & 1188 & 12.87 & 1.23 & 1.15 & 806.67 & 365 & 35.29 & -- & -- & 0 & 0.18 & RS\\
393758688987098240 & ZTF J000918.04+500109.4 & 201783173 & 2.32520 & 50.01928 & 1634 & 15.52 & 1.55 & 1.02 & 504.29 & 476 & 4.33 &  -- &  --& 0 & 0.35 & SSG\\
383280858567817216 & ZTF J001326.17+411113.9 & 194387651 & 3.35907 & 41.18720 & 1579 & 13.97 & 1.19 & 0.94 & 430.34 & 496 & 12.94 & 13.15 & 0.08 & 1 & 0.18 & RS\\
2445149040805181440 & IQ Psc & -- & 3.78163 & -3.33347 & 401 & 11.16 & 1.37 & 1.59 & 1501.03 & 255 & 8.84 & -- & -- & 0 & 0.11 & cut\\
392918215425799552 & ZTF J001512.03+491712.5 & 202040764 & 3.80016 & 49.28681 & 651 & 13.05 & 1.5 & 1.33 & 1171.59 & 595 & 2.24 & 1.12 & 0.06 & 1 & 0.38 & SSG\\
2863714254002754944 & ZTF J001556.74+335317.3 & 365964574 & 3.98645 & 33.88814 & 1350 & 14.8 & 1.34 & 1.09 & 406.95 & 342 & 6.63 & 6.57 & 0.25 & 1 & 0.15 & SSG\\
\enddata

\tablecomments{Source ID, RA, Dec., G, $G_\mathrm{BP}-G_\mathrm{RP}$, distance (inverted parallax), RUWE, $\chi^2_\mathrm{AL}$, and $N_\mathrm{AL}$ come from the \textit{Gaia} EDR3 catalog \citep{GaiaEDR3}. VSX ID and VSX period come from the VSX catalog \citep{Watson2006}. \textit{TESS} periods (P$_\mathrm{TESS}$),\textit{TESS} amplitudes (Amp.), and number of \textit{TESS} sectors used (N$_\mathrm{sec}$) from \textit{TESS} light curves are as described in text, and we provide a cross-reference of the ID from the \textit{TESS} Input Catalog (TIC; \citealt{Stassun2019}). $A_V$ indicates the line-of-sight extinction we use for each source taken from the \texttt{vespa} map. The final column indicates our classification of the source (RS= an RS CVn star; SSG= a sub-subgiant, cut= removed from sample due to astrometric quality cuts. This table is published in its entirety online in a machine-readable format. A portion is shown here for guidance regarding its form and content.}
\end{deluxetable*}
\end{rotatetable}

%\floatpagefraction{1.0} 
%\newpage
%\clearpage

\end{document}